\documentclass[journal]{IEEEtran}
\usepackage{amsmath,amsfonts}
\usepackage[inline]{enumitem}
\usepackage{algorithmic}
\usepackage{algorithm}
\usepackage{array}
\usepackage[caption=false,font=normalsize,labelfont=sf,textfont=sf]{subfig}
\usepackage{textcomp}
\usepackage{stfloats}
\usepackage{url}
\usepackage{verbatim}
\usepackage{graphicx}
\usepackage{cite}
\usepackage{xspace}
\usepackage{xcolor}
\usepackage{cleveref}
\usepackage[normalem]{ulem}

\crefformat{figure}{\figurename~#2#1#3}

\ifdefined\ie\else\newcommand{\ie}{\emph{i.e.}\xspace}\fi
\ifdefined\eg\else\newcommand{\eg}{\emph{e.g.}\xspace}\fi

\title{Towards Deep Application-Network Integration: Architectures, Progress and Opportunities}

\author{
    Berta Serracanta\textsuperscript{*},
\and
    Kai Gao\textsuperscript{*},
\and
    Jordi Ros-Giralt,
\and
    Alberto Rodriguez-Natal,\\
\and
    Luis M. Contreras,
\and
    Richard Yang,
\and
    and Albert Cabellos
\thanks{\textsuperscript{*}Both these authors are the corresponding authors.}
}
\begin{document}

\maketitle

\begin{abstract}
  With the rise of a new generation of applications (e.g., virtual and augmented reality, artificial intelligence, etc) demanding stringent performance requirements, the need for networking solutions and architectures that can enable a higher Quality of Experience (QoE) is becoming increasingly important.
  While jointly optimizing application and network may increase the applications' QoE and simultaneously improve the utilization of network resources, such a paradigm has had limited success in real production networks. However, with the combination of revolutionary trends in (1) compute processing demands, (2) networking capabilities, and (3) sustainable business models, it is high time the community explores the full potential of deeper integration between application and network.

  In this paper, recent trends observed over the past few years are systematically reviewed. These include the paradigm shift in modern communication services towards computing-driven applications, such as on-site AI training, advances in programmable network technologies like Software Defined Networking (SDN), and new business models incentivizing collaboration and cooperation between parties. Following this, successful scenarios that benefit from various forms of deeper network-application integration are reported, highlighting their considerable potential. A unified framework is then introduced, providing an overview of possible architecture paradigms for network-application integration and bringing awareness to existing abstractions, mechanisms, tools, and their potential combinations. The paper concludes with a discussion of several remaining challenges in building practical network-application integrated systems.
\end{abstract}

% Submission guideline:
% https://www.comsoc.org/publications/magazines/ieee-communications-magazine/author-guidelines/manuscript-submission-policy

\vspace{2mm}
\section{Introduction}
\label{sec:intro}

%In the public Internet,
\IEEEPARstart{W}{ithin} the scope of the Internet, applications and networks are fundamentally decoupled. The network is seen by the application as a black-box resource, in most cases without any notion of limit or scarcity. This decoupling has led to the independent evolution of both domains, co-existing in a flexible and scalable architecture that has successfully led to a worldwide network serving a wide plethora of applications and being used by billions of humans every day. Such a decoupled approach has been supported by a continuous scalability process on the network side, so as to accompany the ever-increasing performance requirements on the applications side.

In this longstanding architecture, applications and networks communicate over the well-established standard Socket API.
The resources of the network are typically allocated using a best-effort policy, while transport protocols provide fair use of communication resources, congestion control, and reliable communications. This approach has been driven by the principle of the network being unaware of any special requirement from the application.

However, the emergence of more sophisticated and distributed applications renders this architecture obsolete. Self-driving cars, AR/VR, distributed AI training and inference, Internet-of-Things (IoT), and Industry 4.0 are notable examples of applications with stringent performance requirements that cannot be deployed at scale in the current public Internet best-effort paradigm.
In contrast to the old monolithic application model, the network is not only the way to reach the application, but is now also part of the application itself. This has implications in terms of reliability, performance, and security, which require a deeper integration between the network and the application.
Current network-application integration (NAI) solutions such as the socket API are not expressive enough for applications to specify such requirements.

The sophistication of emerging applications is certainly challenging the prevalent best-effort delivery mode of the Internet.
This issue has been solved in the past by over-provisioning the network resources, which has led to an ever-increasing push for scalability in communication environments. Network operators are now encountering a saturation point where the energy, cost, and complexity requirements necessary for its maintenance, control, and management are becoming increasingly difficult to sustain. Similarly, application providers are encountering scalability issues to sustain the computing platforms demanded by the ever more complex services. A continuous infrastructure scalability growth without any hint for smart traffic handling motivates a constant race for capacity upgrade for both application and network providers, which is not sustainable.

These emerging applications also impose stringent requirements on the compute platforms supporting them, necessitating not only massive computational power, but also demanding low latency and a high degree of locality to efficiently process and analyze the data.  To accommodate these needs, there is a natural trend to shift computational load away from centralized data centers towards more distributed fog and edge platforms. 

This new \emph{compute continuum} (fog, edge, and cloud) has a non-uniform resource distribution, with different components having different processing and connectivity capabilities, as opposed to classical centralized data centers which are usually built to be highly homogeneous.

A better integration between network and application is a very old and well-known problem. Many efforts have been focused on extending the socket API (e.g., \cite{qsockets}). 
However, this is a difficult path forward because this approach can only succeed by achieving worldwide widespread deployment, requiring that the Internet community at large accepts a particular solution to become the new \emph{de-facto} standard for the socket API. The Internet community includes many actors with different (and often conflicting) goals and, thus, such consensus has not been achieved.

A different approach has been followed by the hyperscalers. These companies operate very large computing and communication infrastructures while having full control over the entire stack, thus not requiring to achieve consensus among different actors. In this context, many interesting proposals have emerged. For instance, Aequitas \cite{aequitas} introduced a distributed, sender-driven, admission control mechanism tailored for data center environments. This scheme ensures Remote Procedure Call (RPC)-level service-level objectives (SLOs) by strategically throttling traffic to enhance latency performance within data center networks.

Driven by the requirements of emerging mission-critical applications and the unsustainable cost of over-provisioning, this paper advocates for the timely pursuit of collaborative network and application integration. The main drivers supporting this argument are: (1) recent advances in the network indicating a technology readiness from the communication infrastructure, and (2) new business models offering incentives for competing actors to cooperate towards this integration. Additionally, a novel and unified framework is introduced, providing an overview of plausible architecture paradigms for network-application integration and highlighting existing abstractions, mechanisms, tools, and their combinations. The paper concludes by detailing a list of relevant research challenges that the community needs to address to achieve efficient network and application integration.

\vspace{3mm}
\section{Deeper Network Application Integration:\\Now is the Right Time}
\label{sec:motivation}

\noindent While network-application integration has a great potential to unlock new levels of sophisticated services leveraging on distributed compute performance, it poses several questions: \emph{Is it really necessary now?} \emph{Is it really
possible now?} and \emph{Is it really worth investing now?} This section answers these questions by asserting that: 1) deeper network-application integration is essential to meet emerging computing demands, 2) current networking technology is prepared to enable this integration, and 3) clear benefits have already been reported by industry pioneers.

\subsection{Motivation: Emerging Computing Paradigm}

The last decade has witnessed the rapid evolution of computing
demands for supporting emerging applications such as the Internet-of-things (IoT),
self-driving vehicles, autonomous industry manufacturing, virtual
and augmented reality (VR/AR), and generative artificial intelligence (AI). Many of these
applications pose heterogeneous but yet strict performance requirements, such as latency,
bandwidth, power consumption, and security requirements. As a consequence, the computing paradigm is being extended from cloud computing, where
most computation happens within data centers, to edge computing, where a
fraction of computation is offloaded to edge servers and user devices located closer to the data providers and consumers. In addition, the application building blocks continue to evolve as older software paradigms based on big monolithic architectures are being replaced by composable ``microservice'' approaches. Following the ``cloud native'' trend, applications are now a collection of interconnected microservices, distributed over the cloud, on-prem, and edge locations.

Resource distribution of this new computing paradigm is non-uniform in the sense that
different components have different processing and connectivity capabilities. This is in contrast with data centers, which are built to be highly balanced and homogeneous. For example,
compute capacity can either be deployed by network operators, application providers, or
representatives of the users, which are geometrically unbalanced and form a
hierarchy across multiple administrative domains.

In this context, it is argued that the discovery and management of Quality of Service (QoS) and Quality of Experience (QoE) in such an infrastructure becomes challenging, if not impossible, without deeper integration and collaboration between the network and the application.

\begin{figure*}
  \centering
  \includegraphics[width=0.8\linewidth]{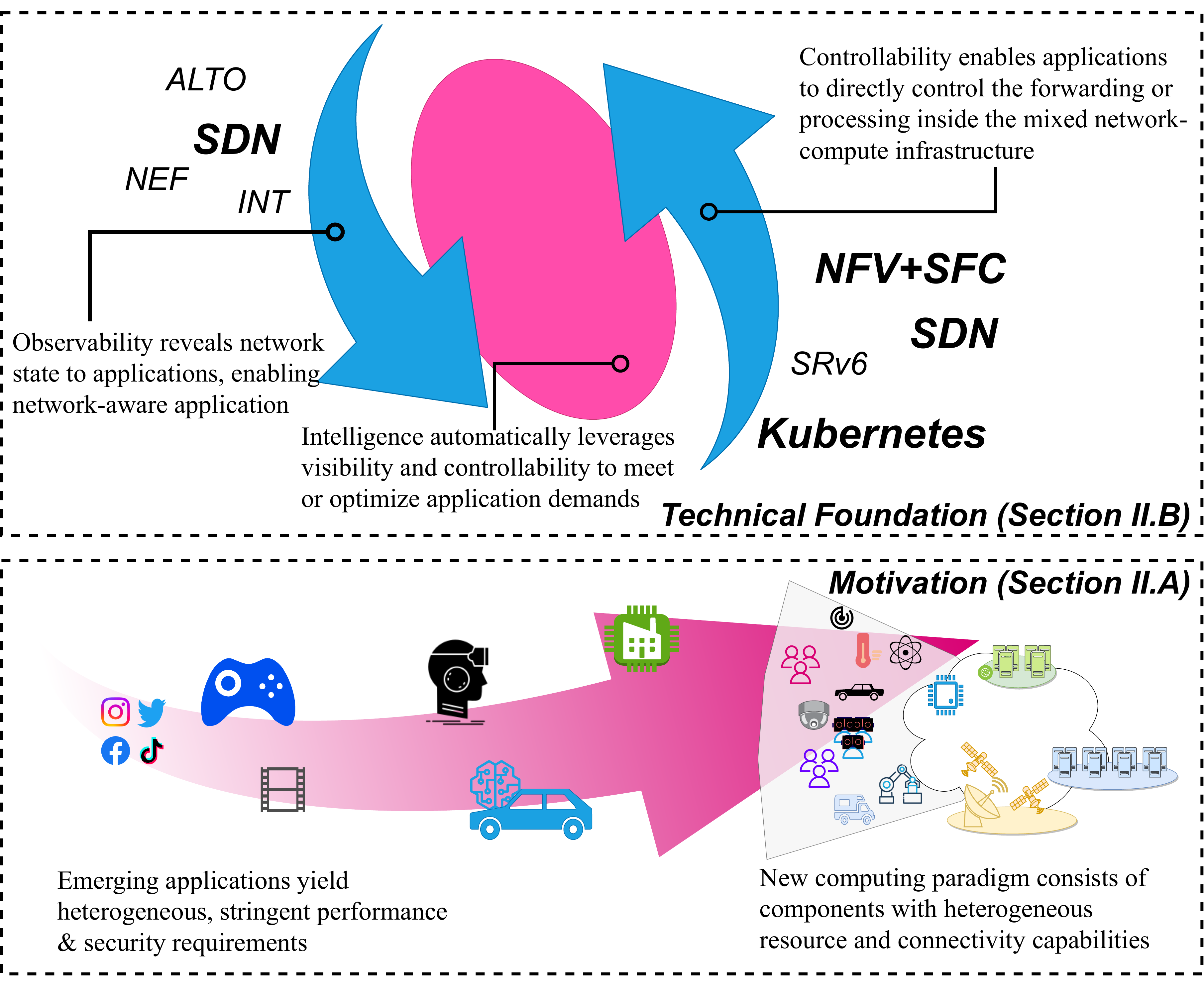}
  \caption{Motivation and technical foundations of deeper network-application integration.}
  \label{fig:timing}
\end{figure*}

\subsection{Foundation: Technology Readiness}

For the past 15 to 20 years, communication networks have evolved to incorporate
a great degree of enhanced capabilities. Through these new technologies,
applications gain better insight of the state of a network (observability) and
have more flexible and automated control of their traffic (controllability),
which together enable more dynamic and optimized application-network integration
(intelligence). Below, some of these key technologies are summarized.

\paragraph{Software-Defined Networking}
Software-defined networking (SDN)~\cite{kreutz2015softwaredefined} has emerged
for more than a decade and has become a representative of the technology trend
of bringing more programmability to the network. The logically centralized SDN
controller can collect states of the entire network to construct a global view,
and conduct coordinated routing control using standard protocols such as
OpenFlow and SRv6. By exposing such capabilities through standard interfaces,
SDN can improve the observability and controllability.

\paragraph{Network Function Virtualization}
Network function virtualization~\cite{nfv} is a technology that
replaces traditional middleboxes with softwarized packet processing logic
(called virtualized network functions) running on general-purpose servers. This
technology has been widely embraced by the industry for its great benefits in
both data center networks and in telco 5G networks, yielding improved
elasticity and control of where and how many network functions are deployed, and reducing costs
of purchasing dedicated middleboxes.

\paragraph{Container-based Computing and Orchestration}
Containers offer a lightweight virtualization technology that substantially
improves the efficiency and agility of virtualized computing. The rise of
container-based computing has led to the great success of Kubernetes~\cite{k8s},
the \emph{de facto} container orchestration system. A Kubernetes cluster is a
great example of network and application integration, which acts as a central
decision point of both resource management and QoE optimization. Kubernetes
may play various roles in the network-application integration framework. For
example, the Kubernetes platform could interact on behalf of the application with the network provider or be used by a network operator
to manage a container-based service edge.

\subsection{Incentives: Benefits, Confirmed!}

Besides the demands increased as new patterns emerge, there are already
successful showcases of deeper network-application integration showing clear
benefits in multiple scenarios, including carrier networks, data center networks,
and cellular networks.

\paragraph{FlowDirector}
FlowDirector is originated from research at
TU Berlin~\cite{flowdirector} and commercialized by Benocs.
The system collects topology and routing information from network service providers and answers queries from communications service providers (CSP) or CDNs through standard protocols such as ALTO~\cite{alto}, to help better determine the mapping from user requests to service points. Benocs reports a 20\% improvement for more than 90\% connections and 15\% reduction in server infrastructure costs.

\paragraph{Aequitas}
Aequitas~\cite{aequitas} is Google's admission control system of Remote Procedure Calls (RPC). The system uses an additive-increase
multiplicative-decrease strategy to adjust the admission probability of RPCs
based on the latency, which serves as an indicator of network resource
contention. Aequitas is able to guarantee 99.9\% RPC network latency (RNL) even
when the traffic is 10x the provisioned demand and reduces an average of 10\%
of the 99.9\% RNL in 50 clusters.

\paragraph{MoWIE}
MoWIE~\cite{mowie} is a joint work by Tencent (application service provider) and China Mobile (network service provider. MoWIE extends Network Exposure Function
(NEF)~\cite{nef}, an interface standardized by 3GPP for mobile cellular
networks, to expose information about the wireless access network collected by
base stations to user equipments. The exposed information is used for better
bandwidth estimation for video streaming and adaptive region-of-interest (ROI)
detection level control for cloud gaming.

\vspace{3mm}
\section{State-of-the-Art}

\noindent In this section, we examine the evolution of the integration between network and applications, focusing on the reasons why past efforts have not seen broad acceptance, despite their appeal. In the second subsection, we discuss more recent efforts and why they have been successful in certain scenarios.

\subsection{Initial efforts}

In the traditional network architecture, TCP and the socket-API are the mechanisms for applications to express their requirements -fundamentally in terms of bandwidth. Networks operate based on a best-effort policy, delaying or dropping packets when resources are unavailable. Beyond such APIs, there has been a plethora of proposals to augment their capabilities.

A notable example of such efforts includes the IETF ALTO protocol \cite{alto}, which facilitates the tuning of network resource usage patterns to boost or maintain application performance. By presenting abstract network maps to the application, the protocol fosters informed decision-making for optimal resource allocation.
QSockets \cite{qsockets}, another instance of socket API extensions, equips applications with the ability to forge and manage connections bearing specific Quality of Service (QoS) requirements. QSockets brings traffic classification on a per-socket basis to the table, offering granular control over packet-specific parameters and facilitating informed scheduling decisions.

Despite the large amount of research efforts devoted to this topic, the widespread deployment of these solutions has proven to be difficult due to different challenges.

First, the current state of network \textit{ossification}. Changing a network's structure is often a complex task, taking considerable time, effort, and resources. This rigidity makes it tough to bring in new features and interfaces. As a result, many networks stick with what they've always done, restricting the adoption of new NAI approaches.

Secondly, the lack of incentives for developers hampers the adoption of NAI solutions. Developers usually adopt new systems when clear benefits like improved functionality or cost savings are evident. While NAI promises enhanced integration, there is a cost and a learning curve associated with these new systems. Furthermore, advanced NAI typically necessitates a change in how developers approach network resources within applications, increasing complexity.

And lastly, cost considerations also affect the deployment of NAI solutions. As the price of computing power went down and cloud-based systems become more common, up until recently it was easier and cheaper for some organizations to simply add more resources when they are dealing with issues related to capacity or performance.

\subsection{Recent efforts}
\label{sec:recent_efforts}
This subsection delves into more recent developments, discussing the success and adoption they have experienced in various scenarios.

First, hypergiant companies own large-scale computing/communicating infrastructures and have control over the entire stack. As a result of such monolithic environment, several NAI solutions have been successfully deployed. 
Notable examples are \cite{aequitas, metaNetworkEntitlement}.
They use specific application information, such as service priority, to develop tailored network mechanisms that guarantee certain quality regions.
In particular, Aequitas \cite{aequitas} leverages both network and application insights to create a distributed admission control mechanism to ensure reliable performance of critical RPCs by constraining the volume of traffic allowed in specific Quality of Service (QoS) regions.
Similarly, Network Entitlement\cite{metaNetworkEntitlement} proposes a large-scale distributed host-based traffic admission system that enforces a contract to achieve network efficiency and meet long-term Service-Level Objectives guarantees on the production traffic.

Second, solutions with limited access and control over specific segments of the stack are often associated with complex environments with different actors and owners. These solutions take a different approach. Instead of explicitly defining an interface between the network and the application, they focus on implicitly inferring the requirements from one party to the other, remaining agnostic to the part of the stack that they cannot control. For instance, some applications in this scenario may have network layer awareness but lack the capability to alter network configurations, while some network services can access application data but cannot control the application layer.

Deep-packet inspection methods constitute a traditional example in this area. These methods, however, often fall short when it comes to identifying encrypted packet streams. This limitation has prompted researchers to explore machine learning-based traffic classification methods. %\cite{ETCNet}.
Alternative approaches like CN-WAN \cite{cnwan} leverage middleware, such as Kubernetes, as intermediaries between the application and the network. By utilizing this middleware, specific application information, such as traffic type or incoming load, can be communicated to the network, enabling proactive adaptation and enhanced resource utilization.

\vspace{3mm}
\section{Architecture Paradigms}
\label{sec:overview}

% Purpose of having an architecture paradigm

\noindent  In this section, we present an architectural view of application-network integration paradigms, with a focus on the mechanisms and abstractions for information exchange between application and network.

\begin{figure*}
  \centering
  \includegraphics[width=0.95\linewidth]{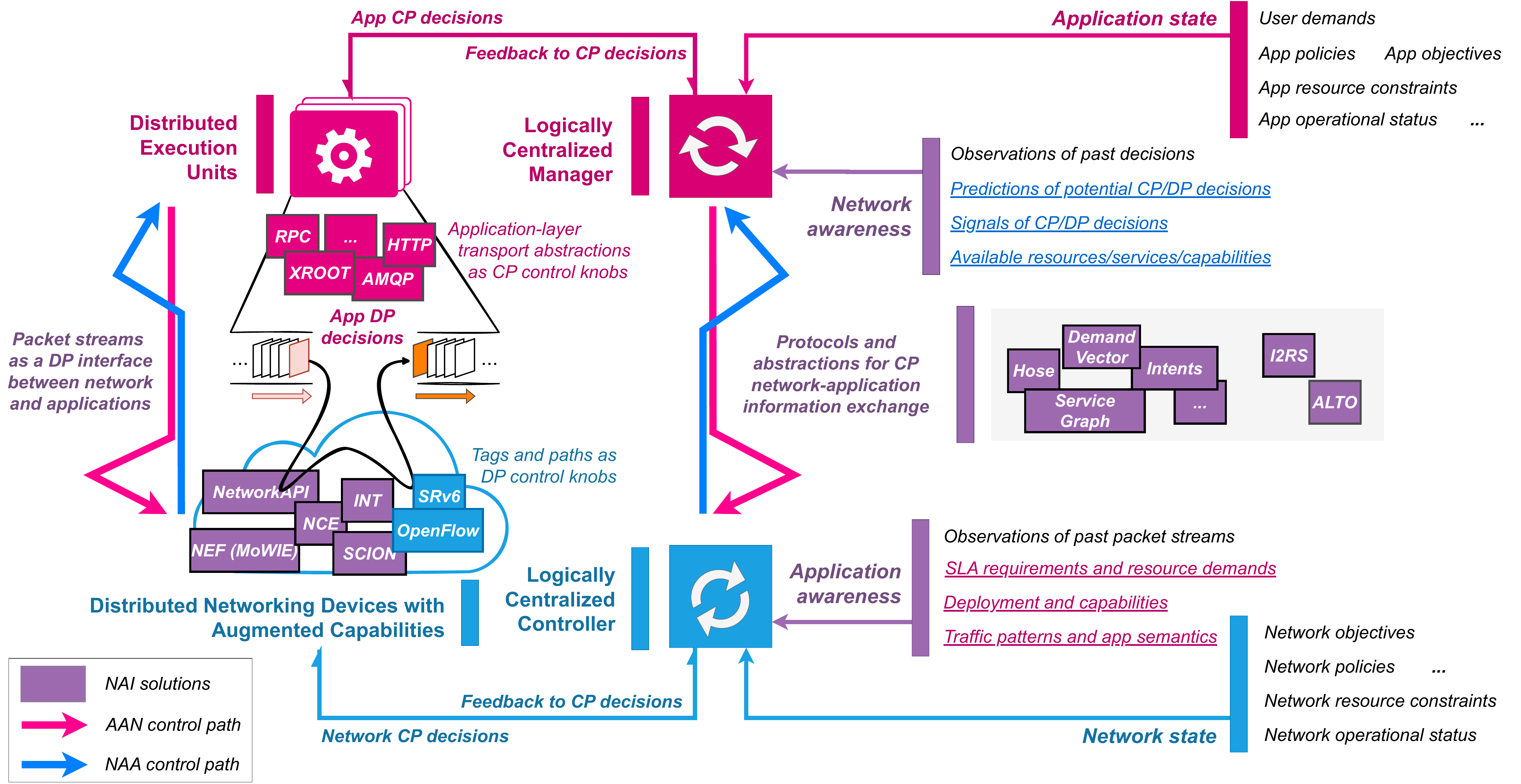}
  \caption{Overview of Potential Architecture Paradigms for Network-Application Integration.}
  \label{fig:overview}
\end{figure*}

% We have a figure to show that the whole control loop of NAI may have two forms: forward control or feedback control
% one paragraph for each component in the loop:
% What does the component mean (definition)?
% What are some common properties?
% 1. application control/decision variables
% 2. application objectives
% 3. network state
% 3.1 static state - irrelevant with the applications' workload/demand
% 3.2 dynamic state - reactive to the applications' workload/demand
% 4. networks' objectives
% 5. applications' state (CATS WG)
% 1. network state abstraction
% 2. forward control
% 3. feedback control

% How do we divide application and network and why?

\subsection{Types of Network-Application Integration}

The proposed framework views the network-application integrated system as a layered architecture, positioning the application at the top and the network at the bottom, as illustrated in \cref{fig:overview}. The application side includes the service provider of a
software logic or service that conducts specific computation tasks to meet its users'
demands, including any software or system components that are involved in the
decision-making of the application's data transmissions. On the other hand, the network side includes
systems and stakeholders that manage and operate multiple resources, either physical or virtual,
providing functionalities such as basic packet forwarding,
encapsulation/decapsulation, as well as other advanced programmable packet
processing logic.

Both layers can be seen as a two-tier system: the \emph{control plane} (CP) and the \textit{data plane} (DP).
The CP of each side is logically
centralized and makes control decisions based on internal states and/or external states of the other side. For example, the CP of the application side conceptually has direct access to the application state (internal) and may gain awareness of the network (external), and vice versa for the network side. The DP consists of multiple distributed components
that carry out the control decisions made by the CP, as shown in \cref{fig:overview}. Note that
the control plane may not necessarily be a real controller but, for instance, a set of coordinated
control components implemented as distributed agents.

With the proposed structure, different network-application solutions can be
categorized by the information exchange tier(s) and the information flow
direction(s).

\paragraph{Information Exchange Tier}
Applications and networks interact in two different ways, via the
\emph{data plane} and via the \emph{control plane} approach. The data plane embeds the exchanged information into packets. Thus,
packet streams provide the most basic abstraction for network-application
integration. Even without explicitly exchanging information, applications and
networks are able to gain awareness of the other side through measurement and
inference. Advanced data plane technologies enable application and network to
inject and interpret specific header fields and/or signals, including metadata, making it possible to
realize semantic-aware packet processing.

The control plane approach, on the other hand, involves additional interfaces or
protocols. The information exchange is between the logically centralized
controllers of both the application and the network with the goal to enable better control decisions.
However, the control plane approach cannot work alone as decisions must be taken
by the distributed application agents or network devices. Examples of information exchanged include
resource requirements (\eg, as in the Network Entitlement case \cite{metaNetworkEntitlement}) or network distances between
end hosts (\eg, as in the ALTO/FlowDirector case\cite{alto}).

\paragraph{Information Flow Direction}
When information flows from the application to the network, as shown by the magenta
arrows in \cref{fig:overview}, it corresponds to an
\emph{application-aware networking} (AAN) solution. In this case, network controllers collect
state from the applications to conduct better traffic engineering or resource
allocation. When information flows from the network to the application, as shown by the blue arrows in \cref{fig:overview}, it corresponds to a
\emph{network-aware application} (NAA) solution. In this case, the application's logically
centralized controller makes decisions with better network awareness.

\subsection{Common Abstractions and Exchanged Information}

While many NAI solutions follow similar control paths as in \cref{fig:overview},
there is a wide spectrum of possibilities on how and what type of information is being exchanged. We now introduce some common options and highlight the scenarios where they apply.

\paragraph{Application abstractions}
Application abstractions are exposed to the network or a third-party traffic
management system, that specifies the needs of the application. Some widely
used abstractions include:

\begin{enumerate*}
\item
  \emph{Service chain/graph} (e.g., \cite{chowdhury2012coflow}) specifies the order or hierarchies of ``function''
  invocations as an annotated graph, with nodes as an execution unit and links
  the relations between the units. It is used by virtualized network functions
  and micro-services, for example to specify the set of flows belonging to the same application job, with the goal to optimize job completion time.
\item
  \emph{Overlay} (e.g., \cite{aequitas}) models the application as a virtual network layer and expresses the
  attributes or demands using a network abstraction, as introduced below. It is
  used by cloud tenants, Internet users, or operators of CDN or VoIP services.
\end{enumerate*}

\paragraph{Network abstractions}
Network abstractions are exposed to developers or operators of applications, or
a third-party application management system, with the ability to specify demands
or query information about the network. It works at a lower layer than
application abstractions. Some common abstractions include:

\begin{enumerate*}
\item
  \emph{Pipe model} (e.g., \cite{qsockets}) abstracts a network as a pipe, a black box with no internal
  structures revealed, that transmit data between two endpoints of an application.
  Pipe models are widely used in today's Internet. Signals such as packet loss,
  latency, throughput, and Explicit Congestion Notification (ECN) are
  foundations of TCP congestion control algorithms, as well as applications
  whose quality of experience depends on the end-to-end network quality of
  service, such as video streaming, web browsing, etc.
\item
  \emph{Hose model} (e.g., \cite{metaNetworkEntitlement}) abstracts a network as a single router with many ``hoses''
  connecting to application endpoints. Unlike the pipe model, a hose does not
  connect two application endpoints but aggregate ingress/egress traffic from/to
  one application endpoint to all other application endpoints through the
  network. It can both be used as an abstraction to specify application demands
  and to get network state, and is used for resource provisioning of Virtual
  Private Networks (VPN) in ISP networks, tenant networks in data centers, or
  distributed data analytics jobs in clusters.
\item
  \emph{Map model} (e.g., ~\cite{alto}) abstracts a network as a matrix where each row represents a
  source application endpoint and each column represents a destination
  application endpoint. Thus, each map contains the network state, such as
  quality of service metrics, control signals, etc., between all source and
  destination pairs. When the desired traffic pattern is sparse, \ie, only a
  subset of (src, dst) pairs are valid, map model can be represented more
  efficiently as a set of (src, dst, metrics) tuples. Map models are used to
  provide quality-of-service metrics to applications such as peer-to-peer data
  transfer, Content Delivery Networks (CDN), or multi-flow scheduling.
\end{enumerate*}

\vspace{3mm}
\section{Challenges and Opportunities}

\noindent The integration of networks and applications remains a significant unsolved problem of the Internet. This paper advocates that the current time is opportune to address this issue, yet the Internet community still faces key research challenges in this field. The following discussion aims to uncover potential solutions and strategies for effective integration.

\subsection{Standard Interfaces}

The definition of standard interfaces, especially at the application layer, has posed challenges in heterogeneous environments with different network administrators. The traditional approach of extending the socket API has proven successful in fully managed stack environments but lacks efficacy in partially operated stack scenarios (as defined in \cref{sec:recent_efforts}). As an alternative, providing incentives to developers could offer a potential path forward to achieve compatibility and interoperability across different application layers.
Another approach that can be taken is Network Tokens \cite{network_tokens}, which facilitate secure coordination between endpoints and networks, giving endpoints explicit control over how their traffic is handled.

While there is no widely deployed solution to extending the socket API, there
are ongoing efforts to get around this challenge through separate channels
(\ie, the arrows between the CP/DP of the application and network side that represent different types of ``control path'' in \cref{fig:overview}).  For example, technologies that offer higher observability include
Application-Layer Traffic Optimization (ALTO)~\cite{alto} and Network Exposure Function
(NEF)~\cite{nef}, which are standards published by the Internet Engineering Task Force (IETF)
and 3rd Generation Partnership Project (3GPP) respectively.

\subsection{Feedback Complexity}

As discussed in section \ref{sec:overview}, network application integration typically requires a feedback signal from the network to the application (or vice-versa). This signal contains information regarding the status of the needs expressed by the application. As an example, the network may signal to the application that it is unable to fulfill the latency requirements. This requires the application to adapt to this signal and change its behaviour. In some scenarios, such as video-streaming, adapting is simple since the workload can be scaled-up or down depending on the needs, for instance by adapting the bit-rate. However, in many other scenarios this adds a new layer of complexity to developers.

\subsection{Scalability}

The widespread adoption of network and application integration in the public Internet means that billions of applications will be sending resource requests to the network in short periods of times. The challenge is, how do you scale this load? To address this challenge, clustering, building hierarchies, and implementing distributed architectures are potential solutions. By organizing network resources and optimizing their utilization, scalability can be achieved, ensuring efficient handling of billions of application requests while maintaining high-performance.

\subsection{The Control Dilemma}
%How much control should the application have over the network?

Merging the domains of application and networking calls for a clear outline of who controls the combined computer-communication system. Some recent work \cite{aequitas} refers to this as \textit{Application-Defined Networking}, representing a new way focused on applications to build NAI systems. However, this approach may not always be suitable in every situation, in cases needing more application intent rather than control. For instance, in cloud computing scenarios, applications often hand over control to cloud managers.

This dialogue is central to the larger discourse on internet governance and control, having profound implications on net neutrality. It highlights the delicate equilibrium that must be maintained between technological authority and socio-economic factors, drawing attention to the important decisions that shape the Internet.

%\vspace{3mm}
%\section{Conclusions}
%{\color{blue}In conclusion, the evolving landscape of network technologies necessitates a more integrated approach between network and application layers. Advances in Software-Defined Networking (SDN) and Network Function Virtualization (NFV) offer a strong foundation for this integration. Emerging business models present new opportunities, promoting collaboration among various stakeholders. However, challenges like standardization, feedback complexity, and scalability remain. Addressing these issues is crucial for improved Quality of Experience (QoE) and resource utilization. A collective effort from both the research and industry communities is essential to harness the full potential of network-application integration for the future of communications technology.}

\vspace{3mm}
\section{Acknowledgments}
\noindent This publication is part of the Spanish I+D+i project TRAINER-A (ref.PID2020-118011GB-C21), funded by MCIN/ AEI/10.13039/501100011033. This work is also partially funded by the Catalan Institution for Research and Advanced Studies (ICREA) and and the Secretariat for Universities and Research of the Ministry of Business and Knowledge of the Government of Catalonia and the European Social Fund.

% MOVED BACKUP TEXT TO SEPARATE FILE backup.tex
 \vspace{5mm}
\bibliographystyle{IEEEtran}
\bibliography{main.bib}

 %\vspace{-8mm}
  \vspace{20mm}
\begin{IEEEbiographynophoto}{Berta Serracanta}
 is a PhD candidate at UPC BarcelonaTech. Her research focuses on network-enabled application acceleration, exploring the integration of network and application layers, and the optimization of distributed systems for enhanced operational efficiency.
 \end{IEEEbiographynophoto}
% \vspace{-8mm}
\begin{IEEEbiographynophoto}{Kai Gao}
 recieved his PhD and BS from the Department of Computer Science and Techonology, Tsinghua University in 2018 and 2012 respectively. He is an associate research scientist at the School of Cyber Science and Engineering, Sichuan University. His research interests include application-network integration, network verification, and programmable control and optimization of distributed systems.
\end{IEEEbiographynophoto}
% \vspace{-8mm}
\begin{IEEEbiographynophoto}{Jordi Ros-Giralt}
is a Director of Engineering at Qualcomm Europe, Inc. He is the author of upwards of 70 papers in academic conferences and journals and more than 25 granted or pending patents. He holds a bachelor's degree in Telecommunications from BarcelonaTech, a MSc and PhD in EECS and an MBA from the University of California.
\end{IEEEbiographynophoto}
 %\vspace{-8mm}
\begin{IEEEbiographynophoto}{Alberto Rodriguez-Natal}
 is a Sr Tech Lead in the Enterprise Networking CTO team at Cisco. He holds a PhD from BarcelonaTech and has coauthored around 15 academic papers, 2 RFCs at the IETF, and 25 granted or pending patents with the U.S. PTO.
\end{IEEEbiographynophoto}
% \vspace{-8mm}
\begin{IEEEbiographynophoto}{Luis M. Contreras}
 (https://lmcontreras.com/) works as Technology Expert at Telefonica CTIO. He holds an M.Sc. in Telecommunications (UPM, 1997), an M. Sc. in Telematics (jointly by UC3M and UPC, 2010), and a Ph.D. cum laude in Telematics (UC3M, 2021).
 \end{IEEEbiographynophoto}
 %\vspace{-8mm}
 \begin{IEEEbiographynophoto}{Y. Richard Yang}
has been a Professor of Computer Science and Electrical Engineering since 2001 at Yale, specializing in computer networks, distributed systems, and mobile computing. His work has influenced 5G technologies, established the ALTO Internet Standard, and received multiple awards and media recognition.
\end{IEEEbiographynophoto}
% \vspace{-8mm}
\begin{IEEEbiographynophoto}{Albert Cabellos}
 (PhD 2008) is a full professor since 2020 at the Computer Architecture Department (Universitat Politècnica de Catalunya). He is the co-founder of the Barcelona Neural Networking (https://bnn.upc.edu/) and the NaNoNetworking Center in Catalunya (https://www.n3cat.upc.edu/).
\end{IEEEbiographynophoto}

\end{document}